\documentclass[sigconf]{acmart}
\usepackage[export]{adjustbox}
\usepackage{wrapfig}
\usepackage{graphicx, subcaption}
\usepackage{lipsum}
\usepackage{amsmath}
\usepackage{array}
\usepackage{multirow}
\usepackage{mathtools}
\usepackage{dblfloatfix}
\usepackage{booktabs}
\usepackage{paralist}
\usepackage{caption}
\usepackage{hyperref}
\AtBeginDocument{%
  \providecommand\BibTeX{{%
    \normalfont B\kern-0.5em{\scshape i\kern-0.25em b}\kern-0.8em\TeX}}}

\acmConference[AIMLSystems '21]{First International Conference on AI-ML Systems}{Oct. 21--23, 2021}{Bangalore, India}
\begin{document}

%%
%% The "title" command has an optional parameter,
%% allowing the author to define a "short title" to be used in page headers.
\title{Resource Constrained Neural Networks for \\5G Direction-of-Arrival Estimation in Micro-controllers}

%%
%% The "author" command and its associated commands are used to define
%% the authors and their affiliations.
%% Of note is the shared affiliation of the first two authors, and the
%% "authornote" and "authornotemark" commands
%% used to denote shared contribution to the research.
% \author{Dr Sumit Darak}
% % \authornote{Both authors contributed equally to this research.}
% \email{sumit@iiitd.ac.in}
% \orcid{1234-5678-9012}
% % \author{G.K.M. Tobin}
% \authornotemark[1]
% \email{webmaster@marysville-ohio.com}
% \affiliation{%
%   \institution{Institute for Clarity in Documentation}
%   \streetaddress{P.O. Box 1212}
%   \city{Dublin}
%   \state{Ohio}
%   \country{USA}
%   \postcode{43017-6221}
% }

% \author{Piyush R. Sahoo, Romesh Rajoria, S. J. Darak}
% \affiliation{%
%   \institution{Algorithms to Architecture Lab, ECE, IIIT Delhi}
% %   \streetaddress{1 Th{\o}rv{\"a}ld Circle}
%   \city{Delhi}
%   \country{India}}
% \email{{piyush17173,romesh17184,sumit}@iiitd.ac.in}

% \author{Shivam Chandhok}
% \affiliation{%
%   \institution{Mohamed bin Zayed University of AI}
%   \city{Abu Dhabi}
%   \country{United Arab Emirates}}
%   \email{shivam.chandhok@mbzuai.ac.ae}
  
%   \author{Danilo Pau, FIEEE}
% \affiliation{%
%   \institution{STMicroelectronics}
%   \city{Milan}
%   \country{Italy}}
%   \email{danilo.pau@st.com}
  
%   \author{Hem Dutt Dabral}
% \affiliation{%
%   \institution{STMicroelectronics}
%   \city{Greater Noida}
%   \country{India}}
%   \email{hem.dabral@st.com}
  
\author{Piyush Sahoo\textsuperscript{1}, Romesh Rajoria\textsuperscript{1}, Shivam Chandhok\textsuperscript{2}, S. J. Darak\textsuperscript{1}, Danilo Pau\textsuperscript{3}, 
Hem-Dutt Dabral\textsuperscript{4}\\
\textsuperscript{1}Algorithms to Architecture Lab, ECE, IIIT Delhi, India \\
\textsuperscript{2}Mohamed bin Zayed University of AI, UAE,\\ \textsuperscript{3}STMicroelectronics, Milan, Italy,\\ \textsuperscript{4}STMicroelectronics, Greater Noida, India.  \\
%{\tt\small \textsuperscript{1}ankan.bhunia@mbzuai.ac.ae }
}

%%
%% By default, the full list of authors will be used in the page
%% headers. Often, this list is too long, and will overlap
%% other information printed in the page headers. This command allows
%% the author to define a more concise list
%% of authors' names for this purpose.
\renewcommand{\shortauthors}{Piyush et al.}

%%
%% The abstract is a short summary of the work to be presented in the
%% article.

\begin{abstract}

With the introduction of shared spectrum sensing and beam-forming based multi-antenna transceivers, 5G networks demand spectrum sensing to identify opportunities in time, frequency, and spatial domains. Narrow beam-forming makes it difficult to have spatial sensing (direction-of-arrival, DoA, estimation) in a centralized manner, and with the evolution of paradigms such as artificial intelligence of Things (AIOT), ultra-reliable low latency communication (URLLC) services and distributed networks, intelligence for edge devices (Edge-AI) is highly desirable. It helps to reduce the data-communication overhead compared to cloud-AI-centric networks and is more secure and free from scalability limitations. However, achieving desired functional accuracy is a challenge on edge devices such as microcontroller units (MCU) due to area, memory, and power constraints. In this work, we propose low complexity neural network-based algorithm for accurate DoA estimation and its efficient mapping on the off-the-self MCUs.  An ad-hoc graphical-user interface (GUI) is developed to configure the STM32 NUCLEO-H743ZI2 MCU with the proposed algorithm and to validate its functionality. The performance of the proposed algorithm is analyzed for different signal-to-noise ratios (SNR), word-length, the number of antennas, and DoA resolution. In-depth experimental results show that it outperforms the conventional statistical spatial sensing approach.

\end{abstract}

%%
%% The code below is generated by the tool at http://dl.acm.org/ccs.cfm.
%% Please copy and paste the code instead of the example below.
%%
\begin{CCSXML}
<ccs2012>
 <concept>
  <concept_id>10010520.10010553.10010562</concept_id>
  <concept_desc>Computer systems organization~Embedded systems</concept_desc>
  <concept_significance>500</concept_significance>
 </concept>
 <concept>
  <concept_id>10010520.10010575.10010755</concept_id>
  <concept_desc>Computer systems organization~Redundancy</concept_desc>
  <concept_significance>300</concept_significance>
 </concept>
 <concept>
  <concept_id>10010520.10010553.10010554</concept_id>
  <concept_desc>Computer systems organization~Robotics</concept_desc>
  <concept_significance>100</concept_significance>
 </concept>
 <concept>
  <concept_id>10003033.10003083.10003095</concept_id>
  <concept_desc>Networks~Network reliability</concept_desc>
  <concept_significance>100</concept_significance>
 </concept>
</ccs2012>
\end{CCSXML}

% \ccsdesc[500]{Computer systems organization~Embedded systems}
% \ccsdesc[300]{Computer systems organization~Redundancy}
% \ccsdesc{Computer systems organization~Robotics}
% \ccsdesc[100]{Networks~Network reliability}

%%
%% Keywords. The author(s) should pick words that accurately describe
%% the work being presented. Separate the keywords with commas.
\keywords{DoA estimation, edge-AI, 5G, neural networks, X-CUBE-AI, micro-controller, STM32}

%% A "teaser" image appears between the author and affiliation
%% information and the body of the document, and typically spans the
%% page.
%%\begin{teaserfigure}
%%\includegraphics[width=\textwidth]{sampleteaser}
%%\caption{Seattle Mariners at Spring Training, 2010.}
%%\Description{Enjoying the baseball game from the third-base
%%seats. Ichiro Suzuki preparing to bat.}
%%\label{fig:teaser}
%%\end{teaserfigure}

%%
%% This command processes the author and affiliation and title
%% information and builds the first part of the formatted document.
\maketitle

\section{Introduction}
A never-ending expansion of network connectivity has led to the evolution of the Internet of things (IoT) capable to route heterogeneous information between people and devices on a planetary scale \cite{IOT1,IOT2,IOT3}. From simple temperature and pressure sensors to smart home and smart industry processing units relying on faster networks, IoT has resulted in exponential growth, offering increased productivity, better security, as well as the availability of a huge amount of business-worth data \cite{IOT1,IOT2,IOT3}. The analysis of such data to get useful trends and patterns is itself a challenging task and recent progress in Artificial Intelligence (AI) has led to promising solutions for exploiting the hidden information and correlations embodied into the data \cite{IOTML1,IOTML2,IOTML3}. Conventional cloud-AI IoT networks perform data processing and inferences at resource-rich data centers. Unfortunately, such a centralized approach suffers from data-communication overhead, high unpredictable latency, low security, privacy, and poor scalability \cite{edgeai1,edgeai2,edgeai3,edgeai4}. Furthermore, data centers are not environmentally friendly, energy parsimonious, and green \cite{DCE}. 

To overcome some of these challenges, the Edge-AI distributed approach is being considered as a solution in which edge devices are featuring intelligence to perform inference either independently or in collaboration with other processing units \cite{edgeai1,edgeai2,edgeai3,edgeai4}. Edge-AI is being actively explored for speech, video, environmental, industry 4.0, and autonomous driving applications. For instance, edge-AI includes assisting devices such as cognitive hearing aids \cite{hearingaid}, optical character recognition \cite{OCR,OCR1}, and smart home appliances such as google home, Alexa, etc.  With the introduction of shared spectrum sensing due to limited spectrum availability, 5G networks demand spectrum sensing to identify spectrum opportunities in time, frequency and spatial domains \cite{5GSS1,5GSS2,5GNR}. The beam-forming based multi-antenna transceivers allow spatially separated users to communicate simultaneously over the common spectrum leading to significant improvement in network throughput \cite{mimo}. However, narrow beam-forming makes it difficult to have spatial sensing (direction-of-arrival, DoA, estimation) in a centralized manner, and hence, intelligence at user devices is highly desirable. Edge-AI is also important to bring next-generation computing paradigms such as artificial intelligence of Things (AIOT), ultra-reliable low latency communication (URLLC) services, smart industry 4.0, and distributed networks to reality \cite{edge5G1,edge5G2}. 

In the last decade, machine and deep learning algorithms have significantly outperformed hand-crafted rule-based and classic machine learning approaches in various application domains. In wireless communications, deep learning algorithms are widely used for spectrum sensing and characterization, channel estimation, and resource allocations \cite{WML1,WML2,WML3}. However, most of their deployment is limited to cloud-AI networks due to the availability of virtually unlimited resources (memory, computational power, etc.). The deployment of these algorithms in Edge-AI networks needs careful design and low power mapping on hardware since achieving the desired target accuracy is a challenge due to area, memory and power-constrained micro-controller units (MCU) usually feature. The work presented in this paper attempts to address these challenges for spatial sensing applications in wireless networks by targeting the highest accuracy with the lowest implementation costs and using the MCU resources parsimoniously.

\begin{figure}[!b]
\vspace{-0.3cm}
    \includegraphics[width=0.915\linewidth]{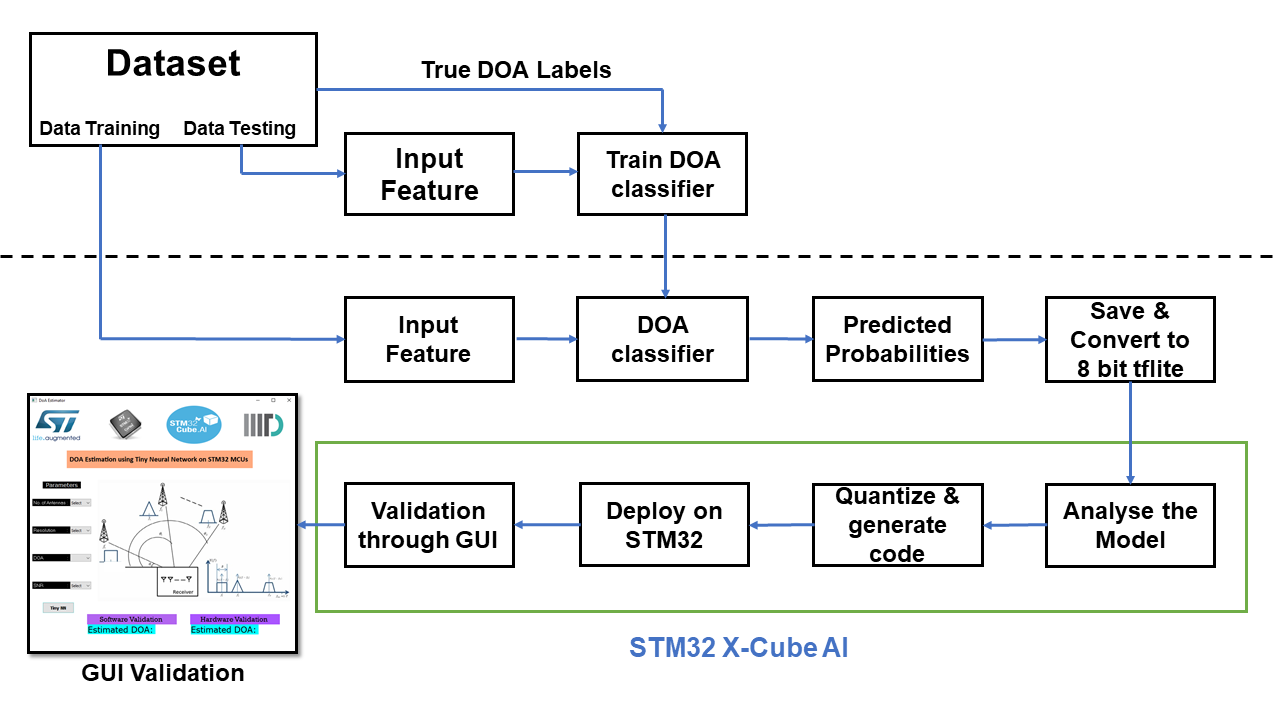}
    \vspace{-0.3cm}
    \caption{Overall architecture for the proposed DoA estimation framework.}% We train our tiny deep learning based DOA classifier on the generated dataset using the supervised learning paradigm. We then deploy it on the micro-controller. By designing an ad-hoc GUI for user interaction, the approach is validated and model performances are reported in real-time. {\color{red} Increase font size in figure and reduce white space (Done)}}
    \label{fig:framework}
    \vspace{-0.3cm}
\end{figure}

\subsection{Problem Formulation}
The goal of this work is to develop a lightweight DoA estimation framework using artificial neural networks that can be easily deployed on off-the-shelf resource-constrained MCUs. Various works have explored DoA estimation for wireless signals and its implementation on hardware such as FPGA/ASICs \cite{doa1,doa2,doa3}. To the best of our knowledge, DoA estimation for wireless signals using neural networks for MCUs has not been explored yet in the literature and is the focus of the proposed work. The DoA estimation-based spatial sensing is also useful in radar, sonar, and navigation to track an object, biomedical to detect tumors, and artery wall movements.

We consider the analog-front-end with multiple antennas and assume that the MCU receives a baseband signal consisting of one active transmission with unknown DoA. The problem of DOA estimation can be formulated as a $K$ class classification problem, where each class corresponds to a possible angle or direction of signal arrival, depending on the antenna-array placement and desired resolution. Given an angle range $A$ of antenna array and a specific DoA resolution $R$, we define a set of DoA classes $C = \{c_{1},c_{2},c_{3} ....c_{K}\}$, where the number of classes, $K=A/R$. In most of the cases, $A=360^{\circ}$. Figure \ref{fig:framework} shows the overall architecture of the proposed DoA estimation framework using a deep-learning-based supervised learning paradigm. Formally, we first generate the training and testing data following the wideband signal generation model described in Section \ref{signalmodel}. We then devise a lightweight neural network architecture and train a deep learning-based DoA classification model as described in Section \ref{architecture}. Finally, we develop a GUI interface for deployment and performance validation as discussed in Section \ref{gui}. Experimental results are discussed in Section~\ref{results} and Section~\ref{conclusion} concludes the paper.

%  Since the hardware we are using here is resource constrained in terms of memory, hence model size is reduced significantly so it can be fit on hardware. 

\section{Proposed Work}
In this section, we discuss the wireless signal model used to generate the dataset and, the design details of the neural network used to estimate the DoA.
\begin{figure}[!b]
\vspace{-0.2cm}
\scalebox{0.8}{
    \includegraphics[width=1\linewidth]{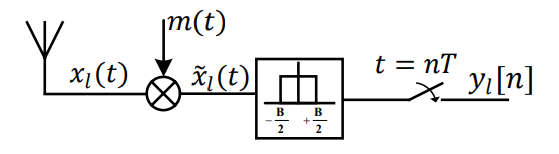}}
    \vspace{-0.2cm}
    \caption{Wideband spectrum digitization approach for dataset generation.}
    \label{fig:antrec}
\end{figure}
\subsection{Wireless Signal Model for Dataset}
\label{signalmodel}
We consider a wideband spectrum consisting of several disjoint, uncorrelated, and far-field signals. These signals are imposed on the $L$-antenna receiver with either Uniform Linear Array (ULA). The signal received at the $l^{\mathrm{th}}$ antenna is defined as follows:
\begin{equation}
\begin{aligned}
x_{l}(t) &=\sum_{m=1}^{M} a_{m}\left(t-\tau_{l}\left(\theta_{m}\right)\right) e^{j 2 \pi f_{m}\left(t-\tau_{l}\left(\theta_{m}\right)\right)}+n_{l}(t)
\end{aligned}
\end{equation}
\begin{equation}
\begin{aligned}
&\approx \sum_{m=1}^{M} a_{m}(t) e^{j 2 \pi f_{m}\left(t-\tau_{1}\left(\theta_{m}\right)\right)}+n_{l}(t)
\end{aligned}
\end{equation}
where M is the number of narrow band signals, $a_{m}(t)$ is the amplitude of $m^{\mathrm{th}}$ narrowband signal of carrier frequency, $f_m$ and DoA, $\theta_m$. Furthermore, $\tau_l(\theta_m)$ is the time delay observed by the $m^{\mathrm{th}}$ signal at the $l^{\mathrm{th}}$ antenna and  $n_l(t)$ is the additive white Gaussian noise (AWGN) at the $l^{\mathrm{th}}$ antenna based on the SNR of the communication channel. The output of the $l^{\mathrm{th}}$ antenna is digitized by means of Nyquist sampling where the received signal, $x_l(t)$ is mixed with a mixing function, $m(t)$ = $\sum_{b \in \beta} \alpha_{l, b} e^{-j 2 \pi(b-1) B t}$ as shown in Fig. \ref{fig:antrec}.  Digitization is performed on the set of non-contiguous frequency bands i.e.  $\beta \in\{1,2, \cdots, N\}$. Here, N is the total number of non-overlapping frequency bands in the wideband spectrum, $\alpha_{l, b}$ is a mixing coefficient of $b^{t h}$ frequency band at $l^{\text {th }}$ antenna carefully chosen from the Gaussian distribution, and B is the bandwidth of a frequency band. Next we define the Fourier transform (FT) of the mixed signal, $\tilde{x}_{l}(t)$, as follows
\begin{equation}
\begin{aligned}
\tilde{X}_{l}(f)=\sum_{m=1}^{M} e^{j 2 \pi f_{m} \tau_{l}\left(\theta_{m}\right)} \sum_{b \in \beta} \alpha_{l, b} A_{m}\left(f-\left(f_{m}-(b-1) B\right)\right)
\end{aligned}
\end{equation}
where $A_{m}(f)$ is the FT of mth signal. The mixed signal is passed through the low pass filter (LPF) of cut-off frequency B/2 as $\tilde{X}_{l}(f)$ covers images over the entire frequency. From the LPF, we get the filtered signal which is digitized at a rate of B Hz. The output of the ADC can be expressed as
\begin{equation}
\begin{aligned}
{Y}_{l}(e^{j 2 \pi f N T})=\sum_{b \in \beta_{busy}}
\alpha_{l, b} A_{b}(f-(f_{b}-(b-1)B))e^{j 2 \pi f_{b}\tau_{l}(\theta_{b})}
\end{aligned}
\end{equation}
where $\beta_{busy}\in\beta$ is a selected set of occupied frequency bands, $A_{b}(f)$ is the FT of the signal in $b^{th}$ frequency band with frequency and DoA of $f_{b}$ $Hz$ and $\theta_{b}$, respectively. The FT of the output of all L antennas in the matrix form is given as
\begin{equation}
\begin{aligned}
\textbf{Y}(f) = \textbf{S}\hspace{1mm}\textbf{Z}(f)
\end{aligned}
\end{equation}
where $\textbf{S}$ is a $L\hspace{1mm}\times\hspace{1mm}M$ steering matrix where $\textbf{S}_{l,m} = e^{j 2 \pi f_{m} \tau{l} (\theta_{m})}$. Note that the steering matrix, $\textbf{S}$ contains two variables, carrier frequency, $f_{m}$ and DoA, $\theta_{m}$ for each of the narrowband signal. The signal received at the antennas is added with AWGN noise based on the SNR value of the channel to generate the final signal corresponding to the combination of DoA and SNR values. In this work, we focus on DoA estimation with known carrier frequency and hence, one frequency band is passed to neural network at a time. The details of dataset generation are given in Section~\ref{results}. 
% \subsection{DOA Estimation as a Classification Problem}
% Our aim is to develop a CNN based supervised learning framework for estimating the DOAs of multiple simultaneously active sources by learning the mapping from the randomly generated narrowband signals to the DOA of these signals using a large set of labeled data. Since the hardware we are using here is resource constrained in terms of memory, hence model size is reduced significantly so it can be fit on hardware. 
% The DOA estimation problem is based on the CNN and it can be formulated as a K class classification, where each class corresponds to a possible angle, which is dependent on the array geometry and the resolution of the whole range. For example, the angle range of an antenna array is from 0 ◦ to 315◦ and the total number of classes is 8, when the resolution is 45◦.

\subsection{Neural Network Architecture}
\label{architecture}
In this section, the proposed neural network architecture for DoA estimation is presented. Specifically, we use a convolution-based neural network (CNN) due to their capability in capturing spatial correlations within input signals at different scales (or receptive field) as demonstrated in various computer vision \cite{alexnet,yolo,rcnn1,resnet,rcnn2,unet,fcn-segmentation} and signal processing \cite{mc1,mc2,mc3,mc4} applications. They are also data-efficient due to the inherent inductive bias of convolution operations \cite{inductive-bias}. Furthermore, CNNs use parameter sharing across inputs, employ operations that are easily both parallelizable and pipelined and entail $O(1)$ sequential operations. This is in contrast to recurrent networks (RNN) which employ sequential processing leading to higher inference time or fully connected networks (FCN), which require more parameters and thus more computational efforts and storage resources when compared with the CNN counterparts \cite{d2l-book,iitkgp}. Furthermore, CNNs are designed to seamlessly work with spatial signals which are in contrast to FCN or RNNs which require additional processing steps to tackle spatial input data. Thus, we choose CNN for building a lightweight memory efficient, therefore deployable on MCU, framework.

% In the previous few decades, Deep Learning has ended up being an integral asset due to its capacity to deal with a lot of information. The interest to utilize covered up layers has beaten conventional procedures, particularly in design acknowledgment. Quite possibly the most well-known deep neural networks is Convolutional Neural Networks (CNN). It utilizes an extraordinary procedure called Convolution. Presently in mathematics, convolution is a mathematical operation on two functions that delivers a third function that expresses how the state of one is modified by the other. It decreases the information into a structure that is simpler to process, without losing features that are critical for getting a decent prediction. Here we have employed this CNN technique to analyze our large dataset and train our model so that it can estimate the DoA of unknown signals accompanied with noise up to the maximum accuracy.
 The proposed CNN model consists of a total of 7 layers. The first two layers are 2D convolutional layers with kernel size 2x2 and Rectified Linear Unit (ReLU) \cite{relu} activation function. Over multiple experiments, we observed that using ReLU allows faster convergence, to learn and perform better. The proposed architecture also includes two 2D max-pooling layers in between which enable progressive reduction of the activations spatial size, reduce the parameters and computational complexity of the network.  Finally, we have two Fully Connected Layers (FC). The first FC layer uses ReLU and the last layer has a softmax function which maps the outputs to probability values for classification. We train our model end-to-end with categorical cross-entropy loss. For training it, we use the Adam optimizer \cite{adam} due to its training stability and fast convergence properties. Specifically, we use a learning rate of $0.0001$ and $\beta_1=0.5$, $\beta_2=0.999$ parameters for Adam the optimizer. Our network is implemented using Keras library \cite{keras} with Tensorflow backend \cite{tensorflow}. Due to limited space constraints, we have not reported various CNN configurations explored during this project. The discussion is limited to a model which offered the best performance fitting also memory constraints of the MCU.
 
\begin{figure}[!h]
\vspace{-0.2cm}
\scalebox{0.8}{
    \includegraphics[width=1\linewidth]{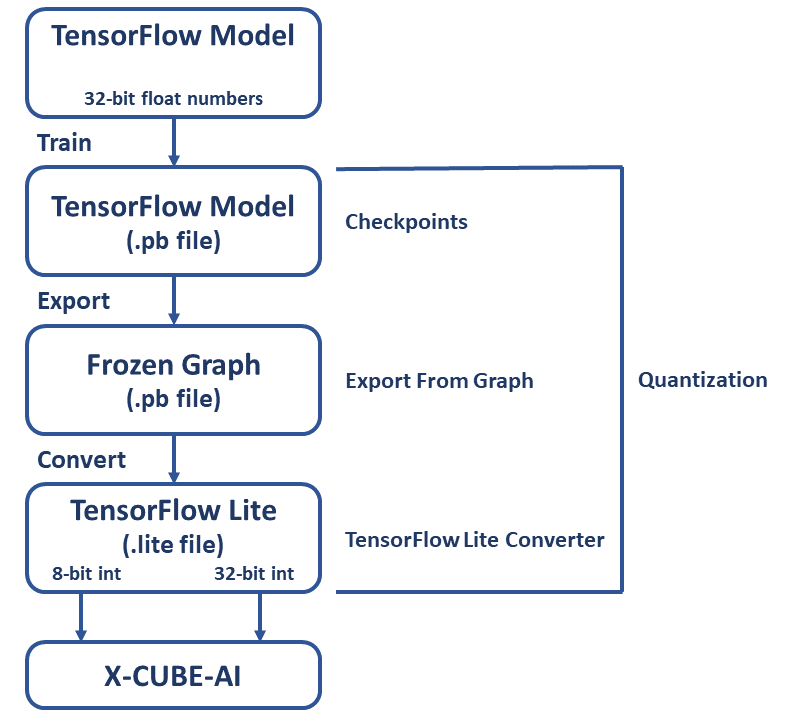}}
    \vspace{-0.2cm}
    \caption{Block diagram depicting the tensorflow steps used to convert proposed CNN model in tflite format that can be deployed on the STM32 micro-controller.}
    \label{fig:compress}
\end{figure}

The next step is to pre-process the model for deployment on the chosen micro-controller and corresponding steps are given in Fig. \ref{fig:compress}. Depending on the application requirements, the model is quantized with appropriate word length (WL). Currently, three WLs, 64-bit double-precision floating-point (DPFL), 32-bit single-precision floating-point (SPFL), and 8-bit integer (INT8) are supported. The quantized model is then exported to as a frozen .pb graph and the file footprint is reduced by discarding metadata and gradient information that were needed during the training procedure. The frozen graph is then converted to the tflite file using the Tensorflow lite conversion procedure. The resulting file is then used to deploy it on the STM32 MCU taking advantage of the X-CUBE-AI automatic deployment procedure integrated into STM32CubeMX \cite{stm32cube}. Please refer to video tutorials (\href{https://www.youtube.com/watch?v=eAHY-xy1N1s&ab_channel=AlgorithmstoArchitectureIIITDelhi}{\color{blue}{demo link}}) explaining the detailing function of the proposed DoA estimation framework.

\section{GUI for Model Deployment on MCU}
\label{gui}
\begin{figure}[!h]
\scalebox{0.95}{
    \includegraphics[width=1\linewidth]{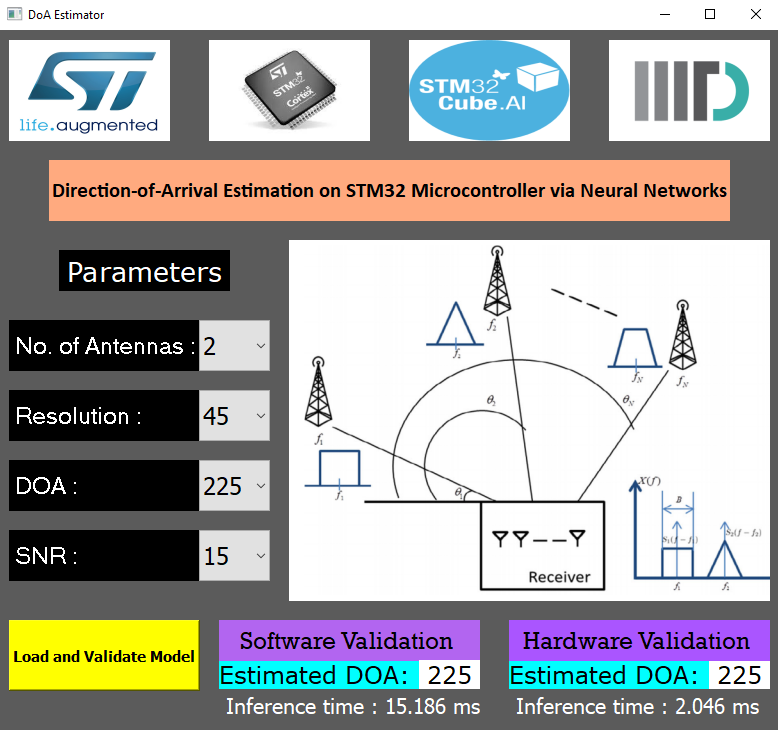}}
    \caption{Proposed GUI to validate the neural-network models in both x86 software and NUCLEO-STM32 hardware modes.}
    \label{fig:gui}
\end{figure}

The deployment of the neural network model on a micro-controller and its performance analysis needs an appropriate user-friendly Graphical User Interface (GUI). Such GUI is application-dependent and needs to be developed from scratch. In this section, we provide details of the proposed GUI which enabled validation of the CNN model on the NUCLEO-H743ZI2 board and comparison with x86 software implementation. 
Fig. \ref{fig:gui} shows the proposed GUI in which users can configure the system parameters such as SNR, number of antennas ($L$), DoA resolution ($R$), and DoA of the active signal along with an appropriate model for a given word-length (WL). The GUI then generates appropriate test data, validates the desired model on the Nucleo-STM32 board, and displays the results given by the software (x86) and hardware (STM32) implementations along with the inference time.

%which is flexible and allows users to generate custom test signals by choosing multiple signal parameters, such as the resolution of DOA, No. of Antennas, DOA class value and SNR value. This GUI is capable of performing combined software and hardware validation in which DOA estimation results for both are show . The two validations are performed simultaneously, one through the model file (software part) and other through the STM32 microcontroller (hardware part). Additionally, in this validation, two models are being deployed on the hardware, one designed for 2 antennas and 45 degree resolution and the other designed for the same number of antennas with 36 degree resolution. They occupy space in RAM as well as in the Flash memory of the board. Hence combined validation is performed for both models in the GUI. 

%\subsection{Hardware validation on STM32 NUCLEO-H743Z12}

To perform the validation on Nucleo-STM32H743ZI2, we develop seamless interaction between the GUI interface and Nucleo-STM32 board to communicate model files and input/output data. In this work, we used X-CUBE-AI-based python libraries to establish the communication between the PC and the NUCLEO-STM32 MCU. There are two python modules composing the HOST API package which play an important role in validating generated neural networks running on a Nucleo-STM32 board connected to the host PC using a standard serial interface (ST-Link V2/V3 virtual com port).

The first module is \emph{stm32nn} which allows discovering the neural networks which are available in the Nucleo-STM32 firmware. It enables GUI application to retrieve important characteristics of the deployed neural network (tensor input/output format, multiply-accumulate operation complexity, RAM/ROM size used by CNN, etc.) and perform inference with the client data (default mode). Furthermore, it also helps to establish the UART communication with the Nucleo-STM32 board. The second module is the \emph{stm32pb} which implements the low-level input/output functions required to communicate with the firmware running on the STM32 device. This firmware communicates with the GUI through the UART interface. For doing Nucleo-STM32 validation we deployed the X-CUBE-AI automatically generated aiValidaiton-application on the edge device (Nucleo-STM32 board)  with the help of STM32CubeIDE which is an Integrated Development Environment for STM32 Microcontrollers. This is a built-in application of X-CUBE-AI that doesn't require any handcrafting to be deployed on Nucleo-STM32.

We used the Nucleo-STM32H743ZI2 board to validate our DoA estimation application. This board features a STM32H743ZI2 MCU (ARM Cortex-M7 core based) which allows IoT practitioners to choose among different combinations of performance and power consumption by selecting among four different CPU run-modes namely Range-0 to Range-3. Run mode Range-0 provides the highest performance at the cost of higher power consumption. Successive run-modes (Range-1, Range-2, and Range-3) can be used to lower the system's power consumption and performance by lowering the CPU operating frequency. STM32H743ZI2 MCU can run at clock frequencies of up to 480MHz and embeds 1 MByte RAM and 2 MBytes of non-volatile flash memory. Due to memory constraints, we need to optimize the size of the CNN model which can be embedded into it. X-CUBE-AI expansion (of STM32CubeMX) package was used to automatically generate STM32-optimized C-code and run time library corresponding to application-specific, pre-trained CNN model. Using X-CUBE-AI, multiple neural network models can be embedded into the STM32H743ZI2 microcontroller provided that the combined size of all the models is within the memory limit of the STM32H743ZI2.
The GUI application also validates the performance of the model on the host processor (x86), which is another feature of X-CUBE-AI. It uses the TensorFlow Lite model file (.tflite) to validate the accuracy of the neural network library generated by X-CUBE-AI. For GUI design, the PyQt5 python library and Qt Designer platform is used for adding different features to make GUI user-friendly. 

\section{Experiments and Performance Analyses}
\label{results}
In this section, we describe the experimental setup and characterize the performances of the proposed framework in comparison with baseline approaches for different values of $R$, $L$, SNR, and WL. We begin with the details of the dataset generated in this project. \\

\noindent \textbf{Proposed Dataset}:
The proposed dataset is generated by using the wideband signal model (described in section \ref{signalmodel}) and MATLAB software. The data generation code requires various input parameters such as $R$, SNR, $L$, sampling frequency, and the number of observations $N$. For the experiments presented in this section, we consider five different DoA resolutions, 5 different size antenna arrays, and 20 different SNR values as shown in Table~\ref{tab:data_param}. Thus, the total number of DoA classes ranges from 8 in $R=45$ to 40 in $R=9$. We  generate \emph{N} $= 2500$ observations for each DoA class. Each observation or data point comprises of complex-valued signal of dimension $l\times d$, where $d$ denotes the dimension of each signal and $l$ is the \emph{antenna count}. The complex-valued data is then converted into the real and imaginary parts. Unfortunately to date, Since Keras does not handle natively complex numbers, they are then concatenated with each other to form a combined matrix of size $(l*2)\times d$, which acts as the input to the neural network.  We keep $80\%$ samples as a training set and $20\%$ for the test set. Furthermore, out of the training set, $20\%$ samples are kept aside as a validation set. All datasets are free and available online (\href{https://drive.google.com/drive/folders/1O-JzVc-f1oNeK9tBtf7zh7SjtwhbUAas?usp=sharing}{\color{blue}{here}}). 
\begin{table}[!h]
    \centering
    \begin{tabular}{|c|c|}
        \hline
         Antenna Count ($L$)& 2-6  \\\hline
         DoA Resolution ($R$)& 45, 36, 24, 18, 9\\\hline
         SNR &0-20dB\\ \hline
         DoA classes &$K= 360/R$\\ \hline
         Observations ($N$) &2500 per class\\
         \hline
    \end{tabular}
    \caption{Summary of signal parameters used for dataset generation}
    \label{tab:data_param}
\end{table}

% \begin{itemize}
%   \item Antennas of range 2 to 6.
%   \item SNR of range 0 to 20 dB.
%   \item DOA resolution: 45,36,24,18,9.
% \end{itemize}
\begin{figure}[!t]
\vspace{-0.2cm}
    \includegraphics[width=1\linewidth]{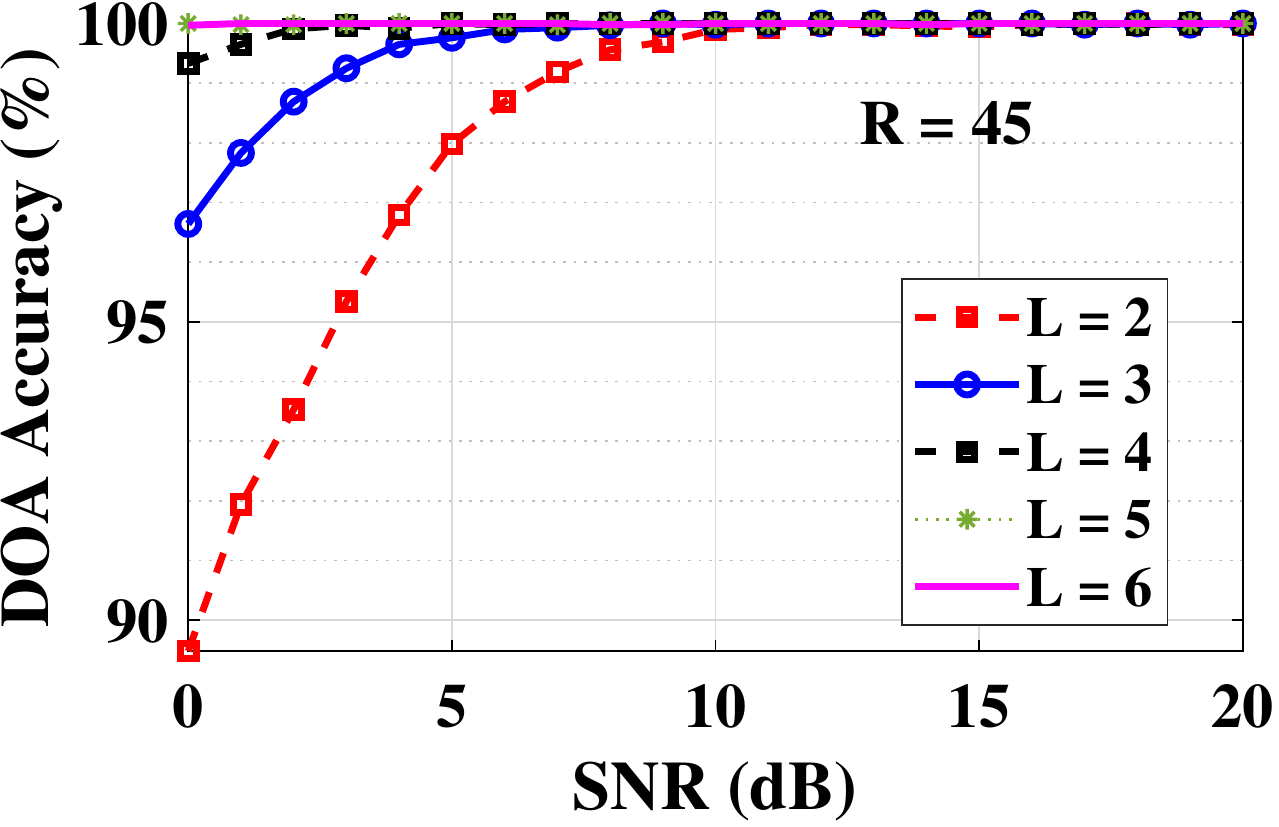}
    \vspace{-0.3cm}
    \caption{Comparison of DoA accuracy in \% for different values of $L$ and SNRs with a fixed $R=45$.}
    \label{fig:fig_1_doavsant45}
\end{figure}
\begin{figure}[!h]
\vspace{-0.2cm}
    \includegraphics[width=1\linewidth]{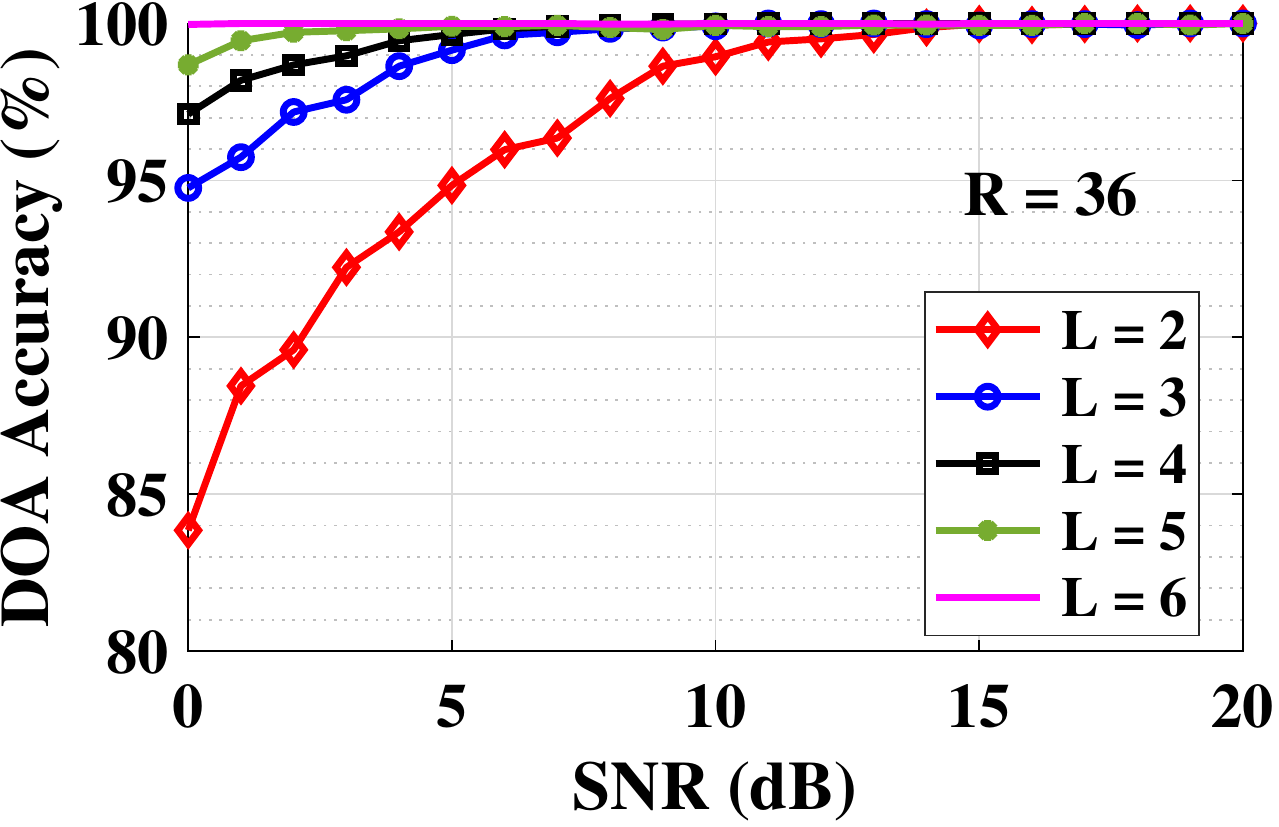}
    \vspace{-0.3cm}
    \caption{Comparison of DoA accuracy in \% for different values of $L$ and SNRs with a fixed $R=36$.}
    \label{fig:fig_1_doavsant36}
\end{figure}

First, we begin with analyzing the effect of $L$ on the DoA estimation accuracy for different SNRs and $R$. In Fig.~\ref{fig:fig_1_doavsant45}, we compare the DoA accuracy for dataset with fixed DoA resolution of 45. Here, we vary the number of antennas from 2-6 and the SNR range is 0-20 dB. It can be observed that the DoA accuracy increases significantly with the increase in the number of antennas due to additional information gained via a signal received from the spatially separated antenna elements. Furthermore, DoA accuracy increases with the increase in SNR due to the reduced impact of noise. Note that the increase in DoA estimation accuracy with the increase in the number of antennas is substantial at low SNR. Furthermore, for the model with only 2 antennas, the average validation accuracy is $96\%$ and the average test accuracy is $95.6 \%$ over the chosen SNR range. In Fig.~\ref{fig:fig_1_doavsant36}, we repeat the same experiments for the dataset with the DoA resolution of 36 i.e. more number of candidate DoAs. The results are similar to Fig.~\ref{fig:fig_1_doavsant45} except for slight degradation in estimation accuracy. From Fig.~\ref{fig:fig_1_doavsant45} and Fig.~\ref{fig:fig_1_doavsant36}, we can conclude that the higher the number of antennas, higher is the accuracy. However, the gain in performance is not significant for the higher number of antennas which means the number of antennas must be selected carefully. This is because an increase in the number of antennas results in an increase in the area and power complexity of the analog front-end due to multiple radio-frequency chains and higher computational complexity of digital-front-end and baseband algorithms due to a higher number of digitized samples.

\begin{figure}[!h]
\vspace{-0.2cm}
    \includegraphics[width=1\linewidth]{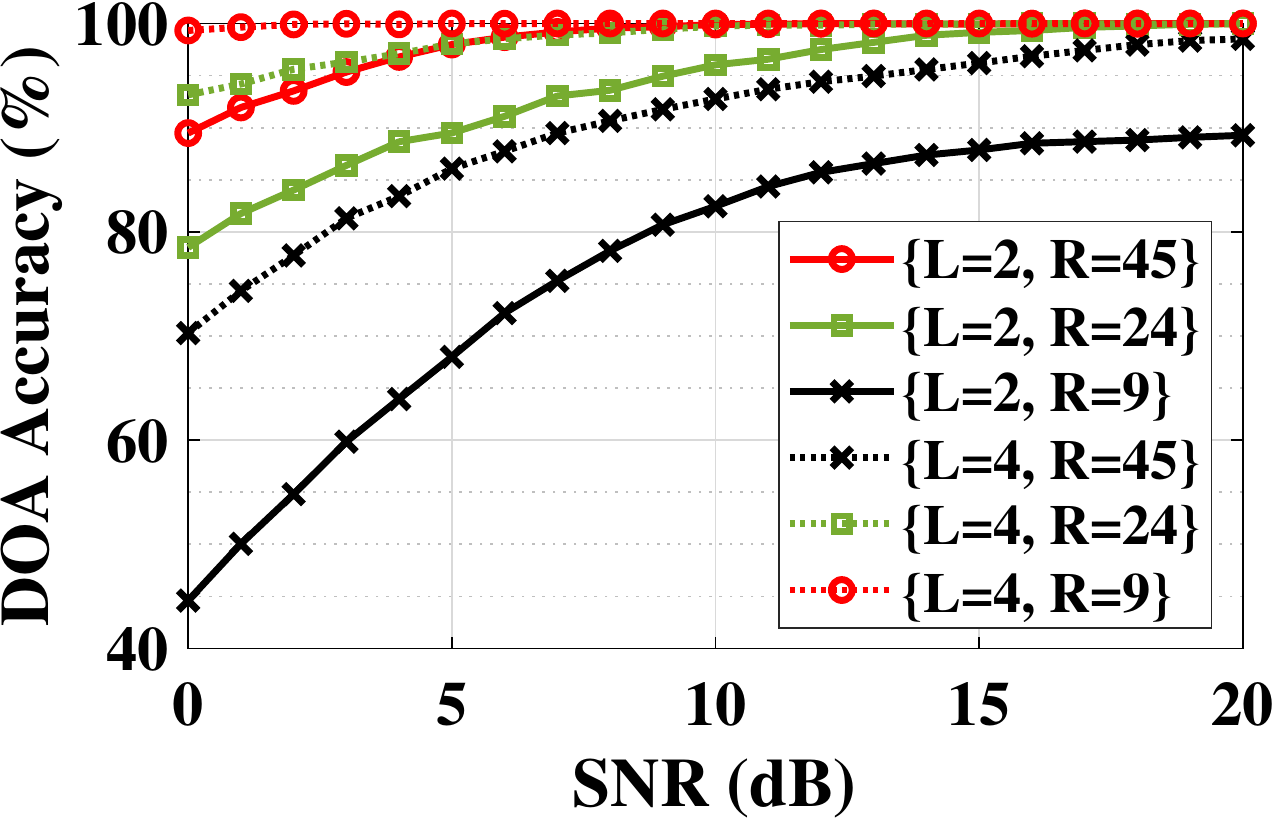}
    \vspace{-0.3cm}
    \caption{Comparison of the effect of $R$ and $L$ on the DoA accuracy in \% for different SNR.}
    \label{fig:fig_3_doavsres}
\end{figure}

Next, we study the difference in the DoA estimation accuracy for different $R$ and subsequent improvement after increasing $L$. In Fig.~\ref{fig:fig_3_doavsres}, it can be observed that the DoA estimation accuracy decreases significantly with the decrease in the DoA resolution. This is due to the increase in the number of classes with the decrease in $R$ making a classification problem more challenging. To further increase the accuracy, we need to increase the number of antennas, especially at low SNRs. This also demands a reconfigurable antenna and DoA estimation architecture that can adapt based on the desired accuracy. Similarly, we study the difference in the DoA estimation accuracy for different $R$ and subsequent improvement after increasing SNR. Since SNR depends on the wireless environment, it can not be controlled and hence, the wireless channel, with the desired SNR, must be chosen via appropriate resource allocation strategies.

 \begin{figure}[!h]
 \vspace{-0.2cm}
    \includegraphics[width=1\linewidth]{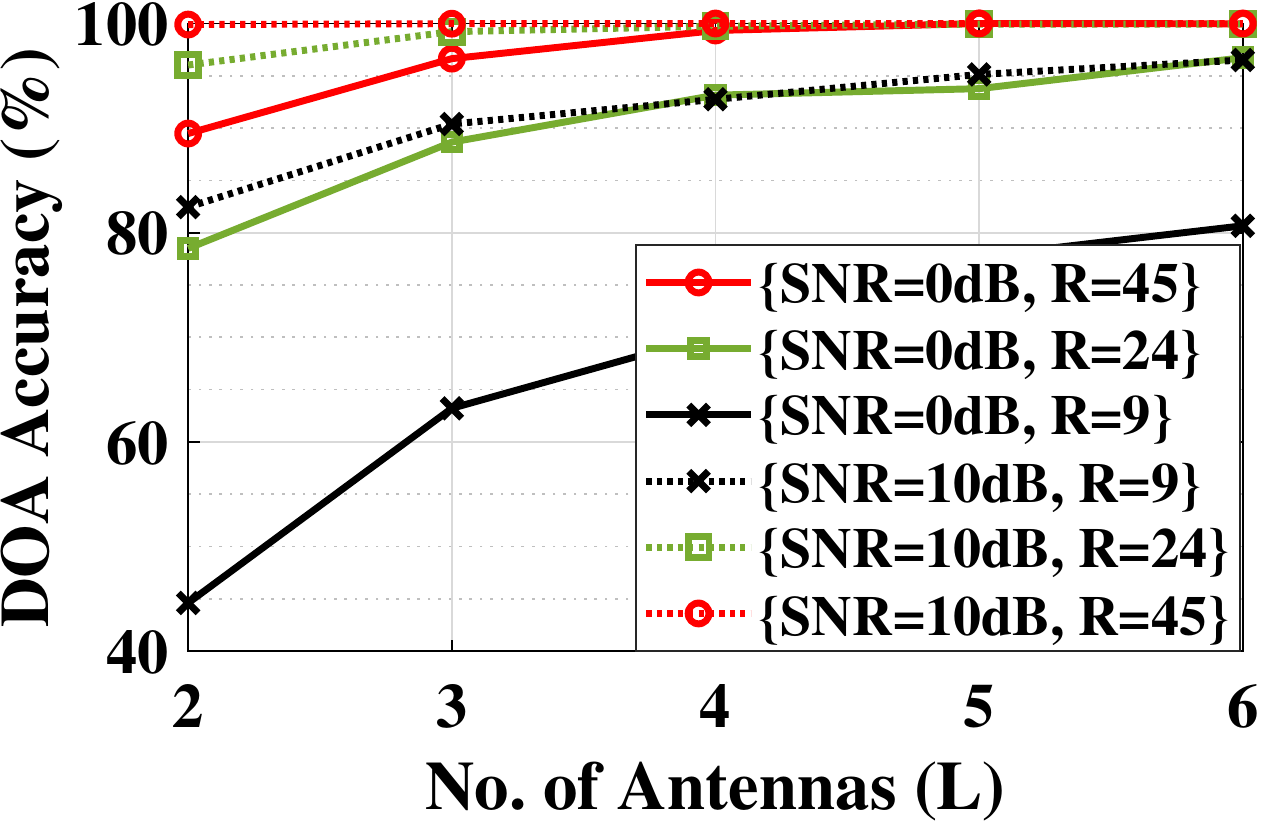}
        \vspace{-0.3cm}
   \caption{Comparison of the effect of $R$ and $SNR$ on the DoA accuracy in \% for different $L$.}
    \label{fig:fig_3_doavssnr}
\end{figure}

\begin{figure}[!h]
        \centering   
        \vspace{-0.2cm}
        \subfloat[]{\includegraphics[scale=0.32]{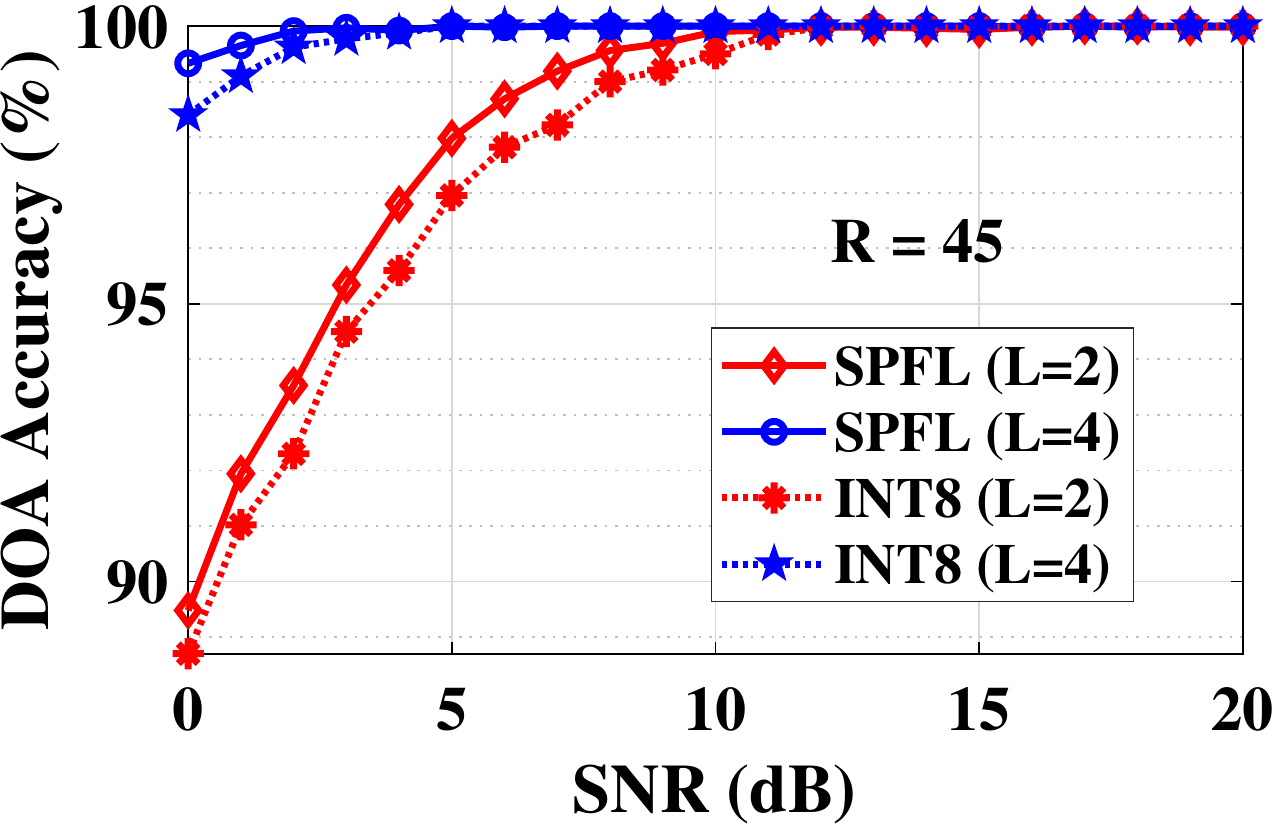}
         \label{fig:quant_1}}
        \subfloat[]{\includegraphics[scale=0.32]{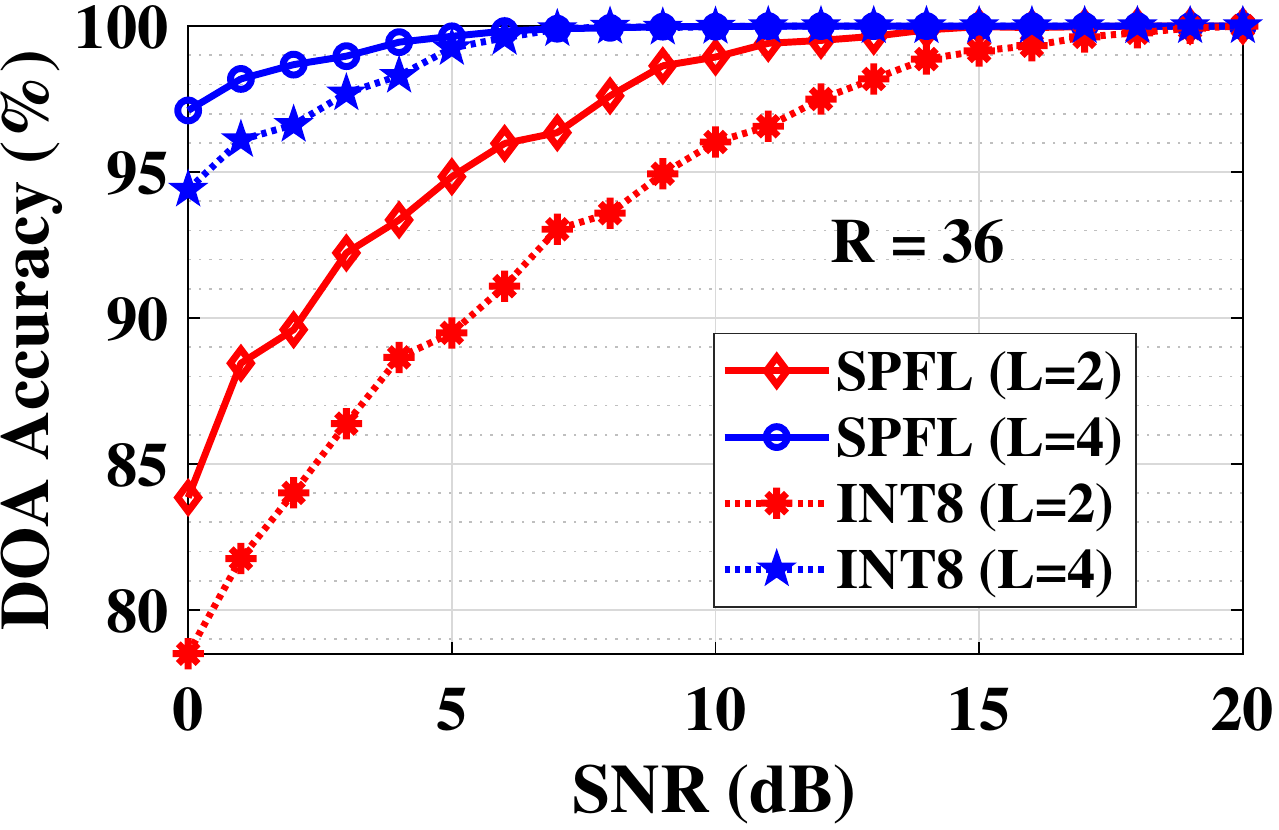}\label{fig:quant_2}}
        \vspace{-0.2cm}
         \caption{Comparison of the effect of WL on the DoA accuracy in \% for (a) $R=45$ and (b) $R=36$ with $L=\{2,4\}$.}
           \vspace{-0.2cm}
\end{figure}

Next, we study the effect of word length on the performance of the proposed algorithm. As shown in Fig.~\ref{fig:quant_1} and Fig.~\ref{fig:quant_2}, WL of 8-bit integer results in slight degradation of the performance compared to SPFL WL. We also observed that the SPFL and DPFL WLs offer nearly identical performance and the DPFL results are excluded to maintain the clarity of the plots. The WL optimization is critical to meet the memory constraints of the micro-controller. For instance, the proposed algorithm with $R=45$, and $L=2$ needs 0.8 Mebibyte (MiB) and 0.3 kibibytes (KiB) flash and on-chip RAM, respectively for SPFL WL. This reduces to 0.2 MiB (-25\%) and 0.09 KiB (-33\%) of flash and on-chip RAM, respectively for INT8 WL. Furthermore, the number of multiply-accumulate computations is reduced from 431560 SPFL operations to 424568 INT8 operations. This has resulted in the reduction in execution time from 6.4 ms to 2 ms (-32\%). Thus, careful selection of WL can offer significant savings in memory, computational complexity and execution time without compromising on the DoA estimation accuracy significantly.

\begin{figure}[!h]
        \centering   
        \vspace{-0.2cm}
        \subfloat[]{\includegraphics[scale=0.32]{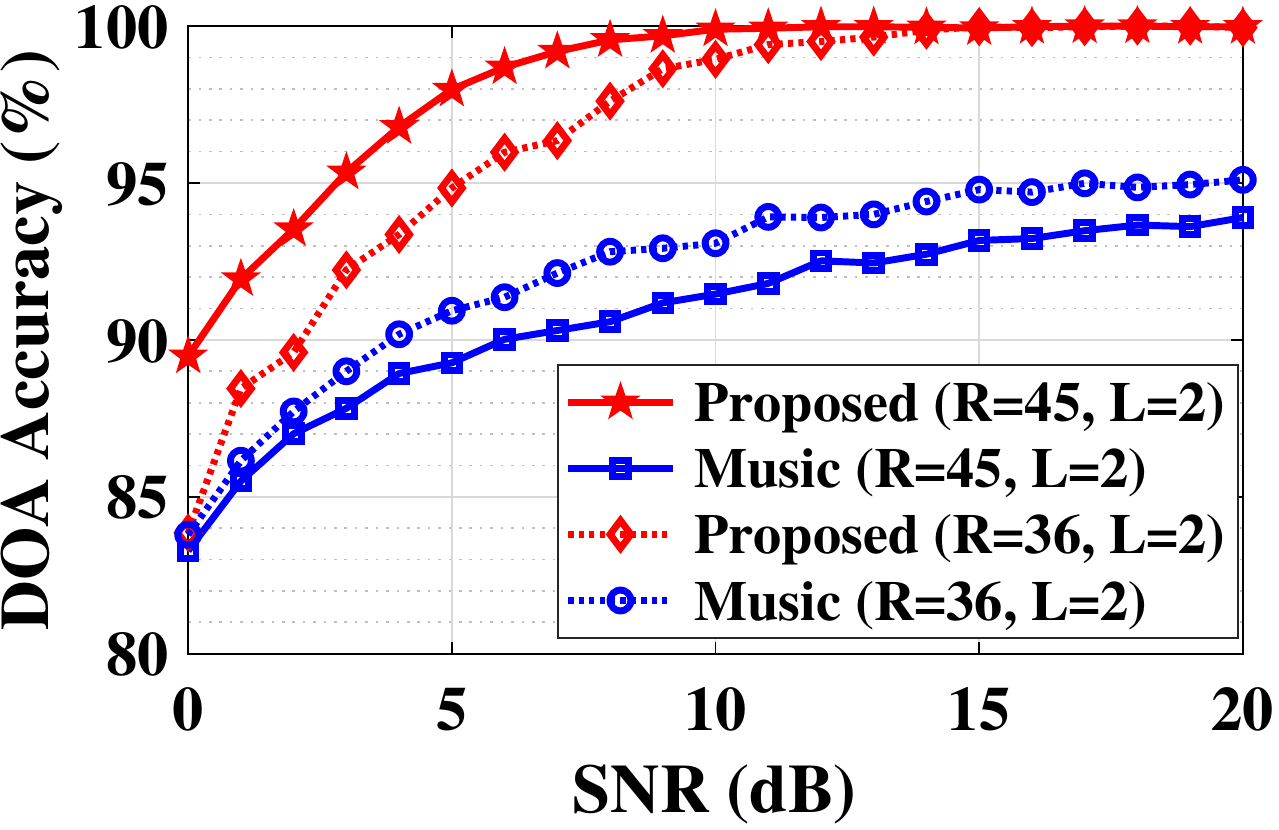}
         \label{fig:music_1}}
        \subfloat[]{\includegraphics[scale=0.32]{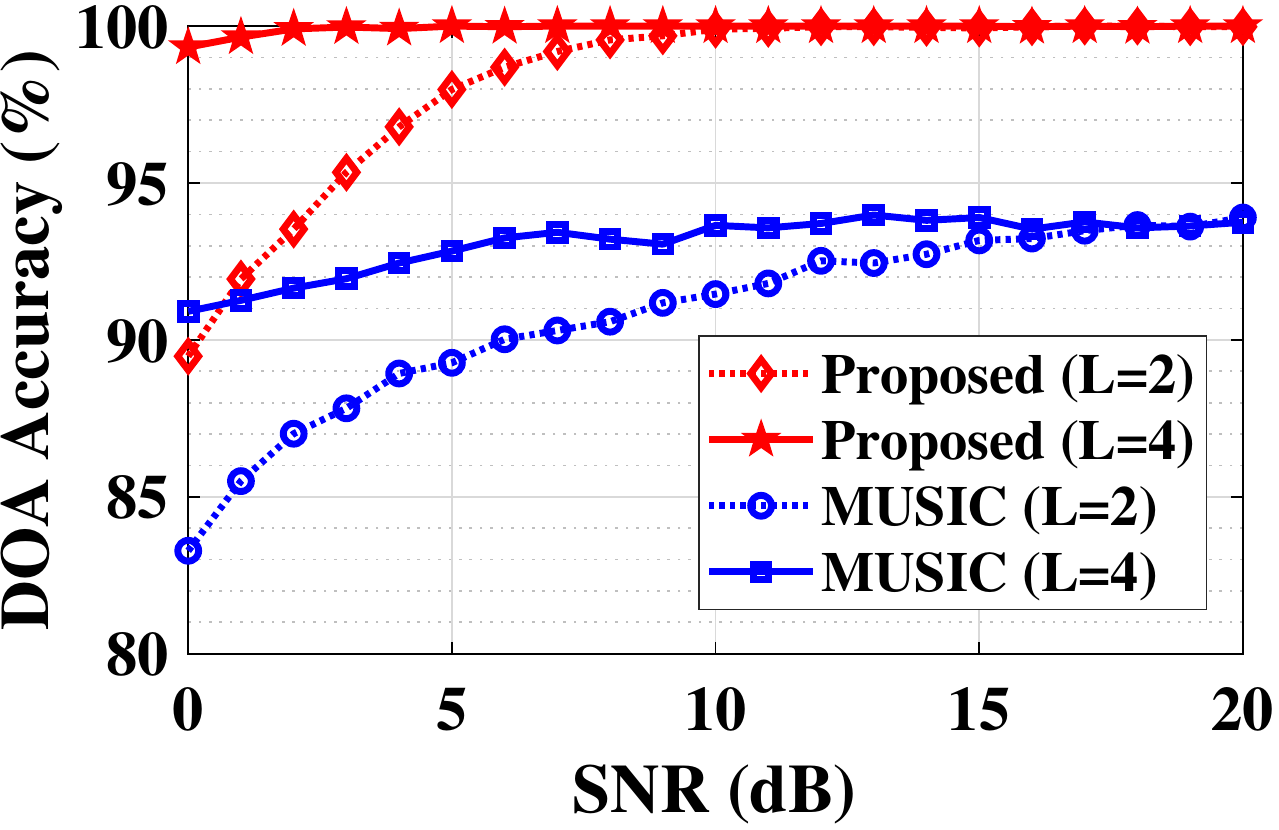}\label{fig:music_2}}
        \vspace{-0.2cm}
         \caption{Comparison of DoA accuracy in \% for MUSIC and proposed approach for: (a) Different $R$, and (b) Different $L$ .}
           \vspace{-0.2cm}
\end{figure}

In the end, we present the comparison of the proposed algorithm with the well-known spatial spectrum estimation approach known as the MUltiple SIgnal Classification (MUSIC) algorithm \cite{music}. The basic idea of the MUSIC algorithm is to perform characteristic decomposition of the covariance matrix of an array output data, resulting in a signal subspace orthogonal with a noise subspace corresponding to the signal components. The two orthogonal subspaces are then used to constitute a spectrum function that can be used for spectral peak search and detect DoA signals \cite{music}. In Fig.~\ref{fig:music_1}, we compare the performance of proposed and MUSIC algorithms for $R=\{45,36\}$ with $L=2$. It can be observed that the proposed algorithm outperforms MUSIC at all SNRs and offers an average of 5\% higher accuracy. As shown in Fig.~\ref{fig:music_2}, the performance of both algorithms improves with the increase in $L$ from 2 to 4. However, the gain in performance is significant in the proposed algorithm compared to MUSIC due to carefully designed CNN architecture exploiting the spatial information received from additional antennas.

% Fig \ref{fig:music-dl} shows the performance comparison of our proposed DoA framework with the baseline MUSIC algorithm for different DoA resolutions. Note the we evaluate the MUSIC algorithm for different tolerance ranges for fair comparison. As expected, the MUSIC algorithm shows better performance with higher tolerance of $\pm 15$ when compared the $\pm 5$ tolerance counterpart. We notice that our proposed DoA model outperforms the best performing MUSIC algorithm variant for both resolutions of $36$ and $45$, across the entire SNR range. Furthermore, we observe that as SNR increases, the performance of our proposed model reaches close to perfect estimation accuracy ($\sim 100\%$ for both the resolutions. In addition to this, we notice that the performance for both MUSIC and our framework increase with increase in SNR.\\
% \textcolor{red}{THERE IS NO COMPARISON ABOUT MUSIC COMPLEXITY VS PROPOSED CNN, EG BY USING NUMBER OF MACC}
% \begin{figure}[!h]
% \vspace{-0.2cm}
%     \includegraphics[width=1\linewidth]{music_1.pdf}
%     \vspace{-0.3cm}
%     \caption{Comparison of DoA accuracy in \% for MUSIC and proposed approach.}
%     \label{fig:music_1}
% \end{figure}
% \begin{figure}[!h]
% \vspace{-0.2cm}
%     \includegraphics[width=1\linewidth]{music_2.pdf}
%     \vspace{-0.3cm}
%     \caption{Comparison of DoA accuracy in \% for MUSIC and proposed approach.}
%     \label{fig:music_2}
% \end{figure}

\section{Conclusion}
\label{conclusion}
In this work, we developed a lightweight, accurate deep learning-based framework for the task of direction-of-arrival (DoA) estimation in wireless networks. We developed a user-friendly graphical user interface (GUI) to deploy and validate the performance of the proposed algorithm on any micro-controller such as the STM32H743ZI2 development board. In-depth performance analysis for different signal-to-noise ratios (SNR), word-length, number of antennas, and DoA resolution validated the superiority of the proposed algorithm over the state-of-the-art approach. Future works include the extension of the proposed framework for multi-band DoA classification and real-time tracking of DoA for upcoming vehicular networks. Furthermore, we would like to incorporate the support for additional word lengths such as INT16 and INT32 which are not available in existing tools. 

\bibliographystyle{ACM-Reference-Format}
\bibliography{main}

%%
%% If your work has an appendix, this is the place to put it.
\appendix

% \section{Research Methods}

% \subsection{Part One}

% Lorem ipsum dolor sit amet, consectetur adipiscing elit. Morbi
% malesuada, quam in pulvinar varius, metus nunc fermentum urna, id
% sollicitudin purus odio sit amet enim. Aliquam ullamcorper eu ipsum
% vel mollis. Curabitur quis dictum nisl. Phasellus vel semper risus, et
% lacinia dolor. Integer ultricies commodo sem nec semper.

% \subsection{Part Two}

% Etiam commodo feugiat nisl pulvinar pellentesque. Etiam auctor sodales
% ligula, non varius nibh pulvinar semper. Suspendisse nec lectus non
% ipsum convallis congue hendrerit vitae sapien. Donec at laoreet
% eros. Vivamus non purus placerat, scelerisque diam eu, cursus
% ante. Etiam aliquam tortor auctor efficitur mattis.

% \section{Online Resources}

% Nam id fermentum dui. Suspendisse sagittis tortor a nulla mollis, in
% pulvinar ex pretium. Sed interdum orci quis metus euismod, et sagittis
% enim maximus. Vestibulum gravida massa ut felis suscipit
% congue. Quisque mattis elit a risus ultrices commodo venenatis eget
% dui. Etiam sagittis eleifend elementum.

% Nam interdum magna at lectus dignissim, ac dignissim lorem
% rhoncus. Maecenas eu arcu ac neque placerat aliquam. Nunc pulvinar
% massa et mattis lacinia.

\end{document}